\documentclass[fleqn,usenatbib]{mnras}

\usepackage{newtxtext,newtxmath}
\usepackage[T1]{fontenc}

\DeclareRobustCommand{\VAN}[3]{#2}
\let\VANthebibliography\thebibliography
\def\thebibliography{\DeclareRobustCommand{\VAN}[3]{##3}\VANthebibliography}

\usepackage{graphicx}	% Including figure files
\usepackage{amsmath}	% Advanced maths commands
\usepackage{physics}
\usepackage{multirow}
\usepackage{orchidlink}
\newcommand{\st}[1]{_\text{#1}}
\newcommand{\ut}[1]{^\text{#1}}
\newcommand{\emax}{E_e^{\mathrm{max}}}
\newcommand{\emin}{E_e^{\mathrm{min}}}
\newcommand{\ecb}{E_e^{\mathrm{cb}}}
\newcommand{\Ee}{E_e}

\newcommand{\Eg}{E_\gamma}
\newcommand{\magic}{MAGIC}
\newcommand{\hess}{H.E.S.S.}
\newcommand{\xrt}{\textit{Swift}-XRT}
\newcommand{\bat}{\textit{Swift}-BAT}
\newcommand{\gbm}{\textit{Fermi}-GBM}
\newcommand{\lat}{\textit{Fermi}-LAT}

\newcommand{\rev}[1]{\textcolor{black}{#1}}

\title[GRB~190114C: a new component?]{Probing the multiwavelength emission scenario of GRB 190114C}

\author[M. Klinger et al.]{
Marc Klinger$^{1}$\thanks{E-mail: marc.klinger@desy.de}\orcidlink{0000-0002-4697-1465},
Donggeun Tak$^{1, 2}$\orcidlink{0000-0002-9852-2469},
Andrew M. Taylor$^{1}$\orcidlink{0000-0001-9473-4758},
and Sylvia J. Zhu$^{1}$\orcidlink{0000-0002-6468-8292}
\\
$^{1}$Deutsches Elektronen-Synchrotron DESY, Platanenallee 6, 15738 Zeuthen, Germany \\
$^{2}$SNU Astronomy Research Center, Seoul National University, 1 Gwanak-rho, Gwanak-gu, Seoul, Korea
}

\date{Accepted 2023 January 6. Received 2023 January 6; in original form 2022 July 15}

\pubyear{2023}

\begin{document}
\label{firstpage}
\pagerange{\pageref{firstpage}--\pageref{lastpage}}
\maketitle

\begin{abstract}
    Multiwavelength observation of the gamma-ray burst, GRB 190114C, opens a new window for studying the emission mechanism of GRB afterglows. 
    Its Very-High-Energy (VHE; $\gtrsim 100$ GeV) detection has motivated an inverse Compton interpretation for the emission, but this has not been tested.
    Here, we revisit the early afterglow emission from 68 to 180 seconds and perform the modeling likelihood analysis with the keV to TeV datasets. We compute for the first time the statistical preference in the combined synchrotron (syn) and synchrotron self-Compton (SSC) model over the syn-only model. In agreement with earlier analyses, between 68 and 110 seconds an unstable preference for the SSC model can be found, which can also be explained by systematic cross calibration effect between the included instruments. We conclude that there is no stable statistical preference for one of the two models. 
\end{abstract}


%%%%%%%%%%%%%%%%%%%%%%%%%%%%%%%%%%%%%%%%%%%%%%%%%%

%%%%%%%%%%%%%%%%% BODY OF PAPER %%%%%%%%%%%%%%%%%%

\section{Introduction}

Catastrophic events such as stellar core collapses can lead to the launching of an ultrarelativistic jet.
The decelerating shock front set up between this jetted outflow and the surrounding circumburst medium redistributes part of the jet's kinetic energy into thermal heating, turbulent magnetic fields, and non-thermal charged particles. The emission produced by the cooling of these non-thermal particles can be observed up to gamma-ray energies and is known as the gamma-ray burst (GRB) afterglow emission.
The earlier time GRB emission, referred to as the prompt emission, shows a complex phenomenology of temporal features. In contrast, the afterglow evolves smoothly, both in time and in its spectral energy distribution. The simplicity of this observable phenomena is consistent with a simple underlying process driving the afterglow emission, providing a clean test-bed to probe the particle acceleration process.

In the standard afterglow scenario, the non-thermal radio to GeV gamma-ray emission is produced by leptonic synchrotron emission, whilst the highest energy gamma-rays are expected to be produced by the inverse Compton upscattering of the synchrotron photons by the parent electron population (synchrotron self-Compton, SSC) \citep{KumarZhang2015}. The recent detections of Very-High-Energy (VHE, $>\!100$ GeV) emission from GRBs \citep{HESS_180720B,MAGIC_detection,HESS_190829A} have offered an opportunity to test this scenario for the very first time.

For GRB~190114C, the VHE emission was interpreted as requiring an additional SSC component \rev{( \cite{MAGIC_newComp}, but also e.g. \citet{WangEtAl2019,DerishevPiran21,GillGranot_review2022})}, although without a statistical significance for this preference being provided. In contrast, the hardness of the VHE emission of GRB~190829A significantly preferred an extension of the synchrotron X-ray emission up to the VHE scale \citep{HESS_190829A}.
This apparent difference in the interpretations for these two GRBs encourages a re-test of the preferred emission scenario for GRB~190114C.

In this paper, we give a statistically robust basis to the question as to whether a preference for an two-component SSC model over a single-component extended synchrotron emission for GRB~190114C is found. We present a full likelihood analysis of the early afterglow of GRB 190114C with the multiwavelength dataset from X-rays to VHE gamma rays. We discuss the afterglow radiation model used here in Section~\ref{sec:model}, followed by the data reduction in Section~\ref{sec:data} and the subsequent fitting in Section~\ref{sec:fitting}. Finally, we present our results in Section~\ref{sec:results}, and discuss the conclusions that can be drawn in Section~\ref{sec:discussion}.

\section{Methods}

\subsection{Reduced afterglow radiation model} \label{sec:model}
We consider a reduced SSC model, within which both the SSC model as well as an extended synchrotron model can be treated as subclasses. 
Our reduced SSC model's photon spectrum consists of a synchrotron and an inverse Compton component. We infer the two components as convolutions of the radiation kernels for synchrotron\footnote{eq. 7.43 of \citet[]{DermerMenon2009}} and inverse Compton scattering \footnote{eq. 2.48 of \citet{BlumenthlGould1970} implemented as in the {\tt AM3} code \citep{AM3_IC_kernel}}, with a non-thermal electron spectrum in a single acceleration and radiation zone. We use two phenomenological parameters to separate between the two models:
Firstly, the parameter $N\st{IC}$ controls the ratio of the peak inverse Compton to synchrotron energy flux. 
Secondly, the parameter $\eta$ controls the electron acceleration rate --- in units of the Bohm acceleration rate --- which, in turn, dictates the maximum synchrotron photon energy. 
Utilizing this parameterization, the parameter range $N\st{IC}>0$ and $\eta > 1$ describes the SSC model, and the parameter range $N\st{IC}=0$ and $\eta < 1$ describes the extended synchrotron model.

The single radiation zone in our reduced SSC model is assumed to be the blast wave behind the forward shock, as commonly defined in the fireball model (see e.g. \cite{KumarZhang2015} for a review and references therein). Several earlier works \citep[e.g.][]{MeszarosRees93,DermerBoetthcerChiang,ZhangMeszaros01,SariEsin2001_FireballSSC,NakarEtAl2009_KNinSSC,PetropoulouMastichiadias09} have explored the dynamics and photon emission (including inverse Compton scattering) of this blast wave. They parameterise multiple assumptions on the magnetic field in this blast wave, the particle acceleration mechanism, and the geometry of the zone using phenomenological parameterizations providing a second, SSC component.

We reduce this phenomenological description further to a level of simplicity needed to answer our question. We therefore merge degeneracies in the overall normalisation into one phenomenological normalisation parameter of the synchrotron photon flux and consider each time bin independently.

Coherently with the fireball model, we define the blast wave as the region behind the relativistic forward shock, which forms between the homogeneous, isotropic plasma shell and a circum-burst medium with constant density, $n\st{up}$. We note the weak dependence of the Lorentz factor on the upstream density $n\st{up}$ (see eq. \ref{eq:LF}), even for non-constant density profiles. When the shell is in its self-similar deceleration phase \citep{BlandfordMcKee1976}, one can estimate the Lorentz factor at the observation time, $t\st{obs}$, from the conservation of the total isotropic energy $E\st{iso}$ as ($m_p$ is the proton mass for a hydrogen dominated circum-burst medium)\footnote{Note the typo in the normalisation factor in eq. 1 of the publisher's version}:
\begin{align}
    \label{eq:LF}
    \Gamma &= \left( \frac{4\pi}{3} m\st{p}c^5 8^3\right)^{-1/8} \left( \frac{E\st{iso}}{n\st{up} t\st{obs}^3} \right)^{1/8}  &\\
    &= 89 \left[ \left(\frac{E\st{iso}}{3.5 \times 10^{53}~{\rm erg}}\right) \left(\frac{1~{\rm cm}^{-3}}{n\st{up}}\right) \qty(\frac{100~{\rm s}}{t\st{obs}})^{3} \right]^{1/8}. & 
\end{align}
We note that $E\st{iso}$ is estimated from the observed gamma-ray flux and a conversion efficiency of kinetic energy to photons needs to be included, which we take here to be 1. We used a value of $E\st{iso} = 3.5\times 10^{53}$ erg, $n\st{up}=1\,\mathrm{cm}^{-3}$ and the geometric means $t\st{obs}=86$s and $t\st{obs}=140$s. We also note that the forward shock is moving slightly faster than the down stream material: $\Gamma\st{shock} = \sqrt{2} \Gamma$.

In order to parameterise the pressures in the downstream region of the shock, we assume that in its rest frame the shock converts a fraction of the incoming ram pressure $\beta^2\Gamma\st{shock}^2n\st{up} m_p c^2$ into thermal pressure ($\varepsilon\st{th}$), outgoing ram pressure ($\varepsilon_\mathrm{ram}$), non-thermal electrons ($\varepsilon_e$) and turbulent magnetic fields ($\varepsilon\st{B}$). The two relative fractions, $\varepsilon_e$ and $\varepsilon_\mathrm{B}$, are assumed to be small enough not to affect the hydrodynamics of the shock. As common within the standard fireball model, the thermal fraction of electrons is neglected in our model \citep[although see e.g.][]{Warren22_thermal}.

From the fraction $\varepsilon\st{B}$ of the incoming ram pressure we estimate the strength $B$ of the homogeneous, turbulent magnetic field as\footnote{We adopted a factor of 32 (fraction of energy density) instead of 16 (fraction of ram pressure) for easier comparability to the typical parameterisation in the literature.}:
\begin{eqnarray}
    B =\sqrt{32\pi \varepsilon\st{B} n\st{up} m_pc^2} \Gamma =  0.6~{\rm G} \left[ \left(\frac{\varepsilon\st{B}}{10^{-3.5}}\right) \left(\frac{n\st{up}}{1~{\rm cm}^{-3}}\right) \right]^{1/2} \frac{\Gamma}{89}. 
\end{eqnarray}
We emphasise that this treatment leaves the magnetic field strength as a free fit parameter, bound by a hard upper limit at ${\varepsilon_B < 1}$. However, it should be noted that values of $\varepsilon_B$ above a few percent already challenge the shock description.

For a given magnetic field strength and system size, the acceleration and cooling timescales can be estimated:
\begin{align}
    &\tau\st{adi}= \frac{2r}{3\Gamma c} = \frac{16}{3}\Gamma t\st{obs} = 4.8\times 10^4 ~ \mathrm{s} ~ \left(\frac{\Gamma}{89}\right) \left( \frac{t\st{obs}}{100~\mathrm{s}} \right)  \label{eq:tadi},&&\\
    &\tau\st{syn} (\Ee) = \frac{9}{8\pi} \frac{h}{\alpha} \qty(\frac{B_\mathrm{c}}{B})^{2} \frac{1}{\Ee} =  10^{3}~ {\rm s} \left( \frac{0.6 ~{\rm G}}{\rm B}\right)^2 \left( \frac{1 \mathrm{TeV}}{\Ee}\right)  \label{eq:tsyn},&&\\
    &\tau\st{acc}(\Ee) = \eta \frac{\Ee}{eBc} = 14~{\rm s} \:\eta \left( \frac{\Ee}{76 \mathrm{TeV}}\right) \left( \frac{0.6 ~{\rm G}}{\rm B}\right) \label{eq:tacc}.&&
\end{align}
Here we used the critical magnetic field strength $B\st{c} = m_e^2c^3/\hbar e = 4.4\times 10^{13}$~G, Planck's and fine structure constant $h=2\pi\hbar$ and $\alpha\st{f}$, and the electron rest mass energy $m\st{e}c^2$. 
As can be seen from fig.~\ref{fig:icCooling}, one can neglect the characteristic timescale of inverse Compton scattering, given that it never dominates compared the other timescales. We explain this in further detail in appendix~\ref{ap:model}.

We assume that the shock continuously, homogeneously, and isotropically injects non-thermal electrons with a power law spectrum $Q \equiv \dd N / \dd \Ee \dd t \propto \Ee^{-p} \exp\qty(-\Ee/\emax)$, above the energy scale $\emin$. The maximum energy scale, $\emax$ is obtained by equating the acceleration and cooling timescales (eq.~\ref{eq:eelmax}). We emphasise that the injection process is poorly understood, and that the commonly used descriptions \citep[e.g.][]{SarietAl96} via time-independent fractions for both the injected particle rate %(via $\zeta_e$) 
and the injected power into these particles %(via $\varepsilon_e$) 
are ad-hoc assumptions. We fix the minimum injected energy to $\emin=10^{9}$ eV.

Equating $\tau\st{adi} = \tau\st{syn}(\ecb)$ and $\tau\st{syn}(\emax) = \tau\st{acc}(\emax)$ yields the two additional energy scales:
\begin{align}
    \ecb =& \frac{9}{4\alpha\st{f}} \qty(\frac{B_\mathrm{c}}{B})^2 \frac{\hbar}{\tau\st{adi}} = 22~\mathrm{GeV} \qty(\frac{0.6 ~\mathrm{G} }{B})^2 \qty(\frac{4.8\times 10^4 ~ \mathrm{s} }{\tau\st{adi}}), &&\\
    \emax =& \qty( \frac{9}{4\alpha\st{f}} \frac{1}{\eta}\frac{B\st{c}}{B})^{1/2} m\st{e}c^2 = 76 ~\mathrm{TeV} \qty(\frac{0.6 ~\mathrm{G} }{B})^{1/2} \qty(\frac{1}{\eta} )^{1/2}. \label{eq:eelmax}
\end{align}

The hierarchy of these three energy scales dictates specific scenarios. The case $\emin \leq \ecb$ ($\emin > \ecb$) is referred to as the slow (fast) cooling regime \citep[e.g.][]{SarietAL98}.

Additionally, we assume that the electron spectrum quickly converges to its time-dependent quasi-steady state distribution \citep[as e.g. in][]{Zhi_Qiu_190829A}, $\dd N / \dd \Ee (t) \propto Q(t) \cdot \tau(t)$, which holds robustly when $\tau $, the shortest timescale, is much smaller than the timescales at which $Q$ and $\tau$ themselves change. Even for ${\tau \sim Q/\dot{Q} \sim \tau/\dot{\tau}}$, the quasi-steady state differs from $Q\tau$ only by a constant factor of order unity. We neglect the corresponding shift of order unity of $\ecb$ in the slow cooling regime.

This spectrum can be phenomenologically described by a smoothly broken power law: 
\begin{equation}
%    \label{eq:qss_electrons}
    \dv{N}{E_{e}} = \eval{\dv{N}{E_{e}}}_{E_{e0}} \qty(\frac{E_{e}}{E_{e0}})^{-p_{1}} \left[1 + \qty(\frac{E_{e}}{E_{e1}})^{s\st{e}} \right]^{ \frac{(p_{1}-p_{2})}{s\st{e}}} e^{- \frac{E_{e}}{\emax}} \theta(E_{e}>E_{e0}).
\end{equation}
For the slow cooling regime, $E_{e0}=\emin$, $E_{e1}=\ecb$, $p_{1}=p$, and $p_{2}=p+1$. On the other hand, for the fast cooling regime, $E_{e0}=\ecb$, $E_{e1}=\emin$, $p_{1}=2$, and $p_{2}=p+1$.
Due to the preference of the smoothness parameter $s\st{e}>10$ found in \cite{FermiNSwift}, we use a value of $s\st{e}=20$, which effectively yields a broken power law for the electron spectrum.

We reduce all the complexity of the normalisation of the two components of the photon spectrum to two parameters: 1) the normalisation of the synchrotron component $F\st{syn}$ at an observed energy of 100~keV and 2) the relative factor of the inverse Compton component $N\st{IC}$, phenomenologically defined as the ratio of the maximum value of the energy flux spectra of both components. We emphasise the flexibility of this approach to model a standard one-zone SSC spectrum as well as different scenarios leading to a single extended synchrotron component above the so called synchrotron burn-off limit of about 100 MeV in the emission frame.

This source spectrum is then transformed to the observer's frame by assuming a point-like emission zone, an observation angle of $0^{\circ}$ and a redshift of $z=0.43$. Since we observe a strong suppression of the photon flux below a few keV we include a significant photoelectric absorption at the source and leave the column density NH as a free parameter in our fit. Additionally, a local absorption in our Milky Way is included \citep[$N_H = 7.54 \times 10^{19} \mathrm{cm}^{-2}$, ][]{FermiNSwift}. We assume throughout that internal $\gamma\gamma$-absorption or secondary cascading lead to only secondary effects, and are therefore safely negligible.

\subsection{Data reduction}
\label{sec:data}

GRB 190114C triggered the \textit{Fermi} Gamma-Ray Burst Monitor ({\gbm}) at 20:57:02.63~UT ($T_{0, \mathrm{GBM}}$) and 0.56s later the \textit{Swift} Burst Alert Telescope ({\bat}) ($T_{0, \mathrm{BAT}}$) \citep{FermiNSwift}.

Its afterglow emission ($\gtrsim$ 30s) was detected by many follow-up observations, in particular the \textit{Swift} X-ray Telescope ({\xrt}), the \textit{Fermi} Large Area Telescope ({\lat}) and the Major Atmospheric Gamma Imaging Cherenkov (\magic) telescopes \citep[][]{MAGIC_newComp}, providing an unprecedentedly broad dataset from keV to TeV energies. 

We revisit two emission phases ($67.71-110$s and $110-180$s after $T_{0, \mathrm{BAT}}$) and perform the joint-fit spectral analysis with multi-wavelength data: {\xrt}, {\bat}, {\gbm}, {\lat}, and {\magic}. 

As a departure point to explore the stability of the model fit comparison, we adopt a similar analysis for the \textit{Swift} and \textit{Fermi} detectors as was used in \citet{FermiNSwift} (our \textit{default} case) and refer to the appendix~B for more details, including the few {\lat} photons detected (Fig.~\ref{fig:LATphotons}). 
Note that this includes a floating-normalization factor \citep[$< \pm$ 15\%, see][]{FermiNSwift} to {\bat} due to the overlapping energy range with {\gbm}. 

For the stability analysis we additionally define a \textit{floating norm}~15\% setup, where we include for each detector a floating norm limited to $\pm 15\%$ to account for the uncertainty in the cross-calibration between instruments \citep{Madsen2016,Tsujimoto2010,BAT_sys,GBM_sys,LAT_sys,MAGIC_moon}.  
 
For the {\magic} data, we do not have access to the raw data, preventing any studies of systematic effects. Instead, we use the publicly available energy flux points in \cite{MAGIC_newComp} and interpret the uncertainties as Gaussian standard deviations.

\subsection{Joint-fit spectral analysis procedure}
\label{sec:fitting}

We fit the model (see Section~\ref{sec:model}) to the data at the counts level by forward-folding the model flux through the instrument response and by using the proper statistical treatment of signal and background to calculate the likelihood for each instrument: {\tt C-stat}\footnote{Poisson data with Poisson background \citep{cash1979}} for {\xrt}, {\tt $\chi^2$} for {\bat} and {\magic}, and {\tt PG-stat}\footnote{Poisson data with Gaussian background} for {\gbm} and {\lat} (see appendix~B for details). Fig.~\ref{fig:countsplot} visualises the unabsorbed, best fit in energy flux space (top) for the \textit{default} SSC case with the counts (rate) of observation, background and expectation and residual for each included detector. 

Following a Bayesian approach, we derive the posterior probabilities of the parameters and integrate the evidence $Z$ for each model (see appendix~C for details). We perform quantitative model comparison via the Bayes factor $B_{12} = Z_1/Z_2$, stating the preference of model 1 over model 2. 
The interpretation of the Bayes factor requires a scale, and we emphasize the potential bias from this choice. A common choice is the Jeffrey's scale, which interprets values for $\log_{10} B_{12}$ of 1 and 2 as thresholds for moderate and strong preference, respectively \citep{Trotta2008}. It should be used carefully since the idea of unifying the interpretation of significance across different branches of science can also be seen as a limitation of its applicability.

\begin{figure*}
    \includegraphics[width=\linewidth]{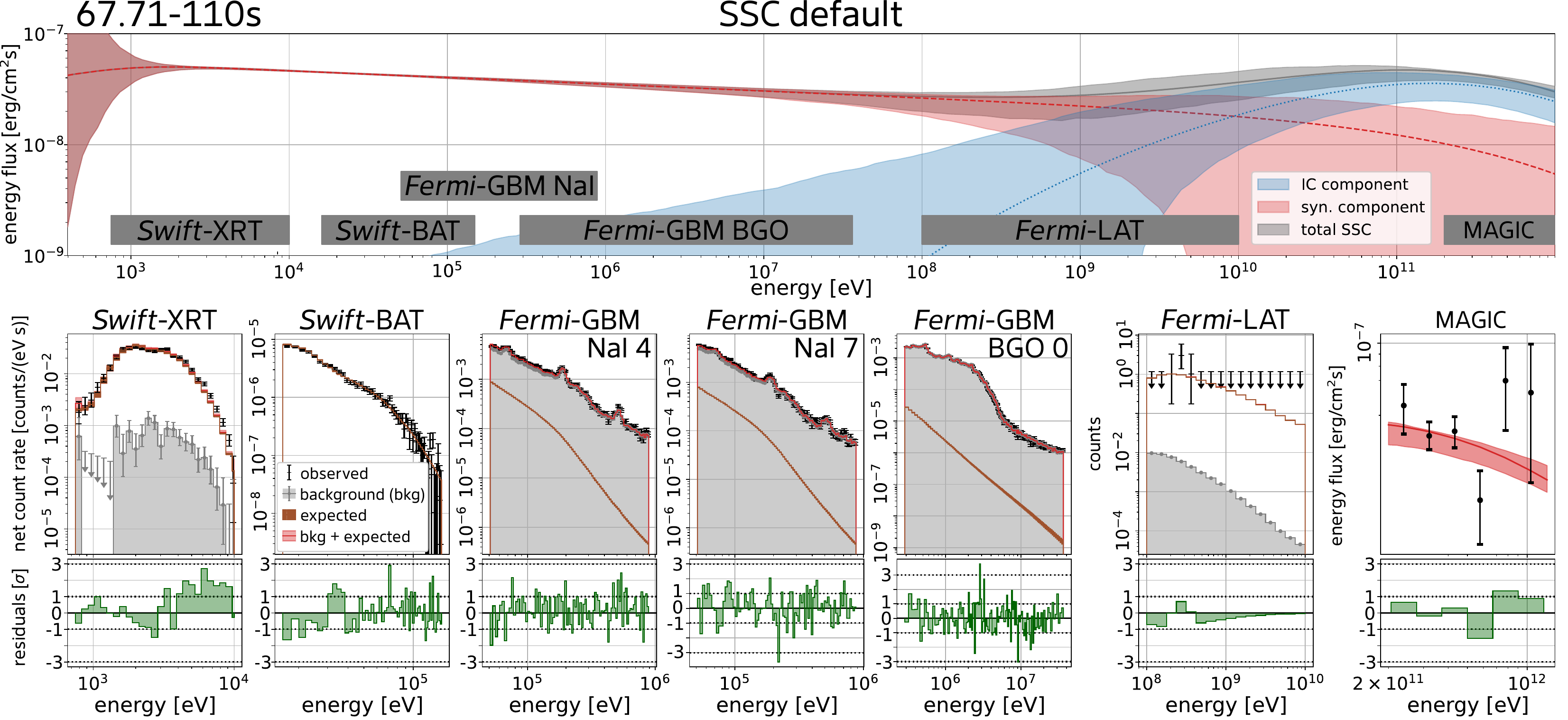}
    \caption{Visualisation of our method. Above: unabsorbed energy flux with the 1-$\sigma$ envelope with its two (red and blue) components. Below: Measured and forward folded model count rates and corresponding residuals for each detector.} \label{fig:countsplot}
\end{figure*}

\section{Results}
\label{sec:results}

\begin{figure*}
    \centering
    \includegraphics[width=0.8\linewidth]{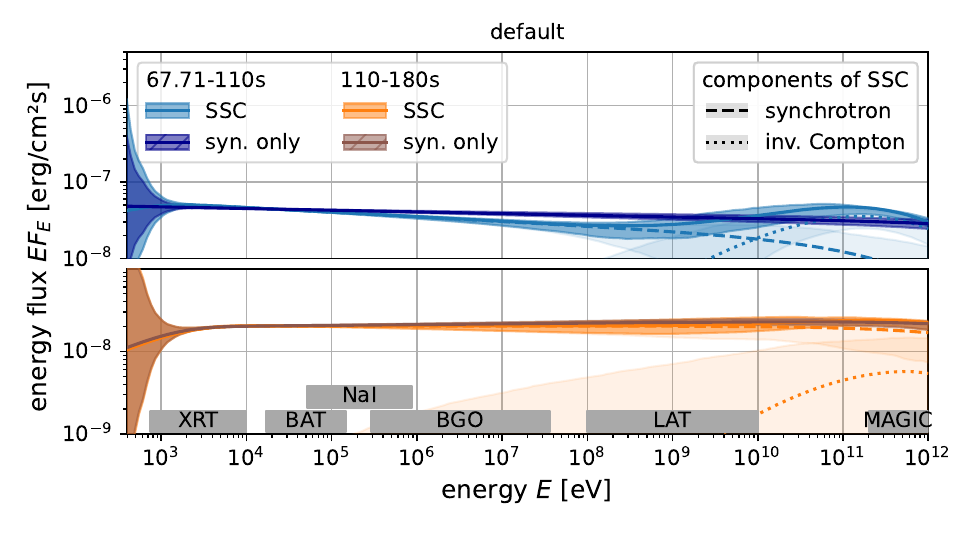}
    \caption{Spectral energy distributions (SEDs) for two time bins for the two models (68\% envelopes with best fit lines) for the default case (see text). The dashed and dotted lines show the best fit components of the SSC model. The grey boxes on the bottom show the energy ranges of the different instruments, which are included in the fit. Note that {\gbm} consists of NaI and BGO detectors (see appendix~B for details).}
    \label{fig:SEDs_default}
\end{figure*}

We show in Fig.~\ref{fig:SEDs_default} the uncertainty bands of the energy flux, $EF_E$, for the time bins (67.71-110~s and 110-180~s after $T_{0,\mathrm{BAT}}$) for the two models (SSC and extended synchrotron only model). As appreciated from the figure, the data require that all four model envelopes are extremely flat, $|\Delta \log_{10}(E F_E) / \Delta \log_{10} E| < 0.5 / 9 $.
Fig.~\ref{fig:SEDs_default} indicates that in the second time bin both models are consistent within their uncertainties, whilst for the first time bin the SSC model differs from the synchrotron only model by more than 1$\sigma$. However, only a statistical test, such as the Bayesian model comparison introduced in Section~\ref{sec:fitting}, allows for any further statement about a preference.

\begin{figure*}
    \centering
    \includegraphics[width=0.85\linewidth, trim={0 0 0 1.5cm}, clip]{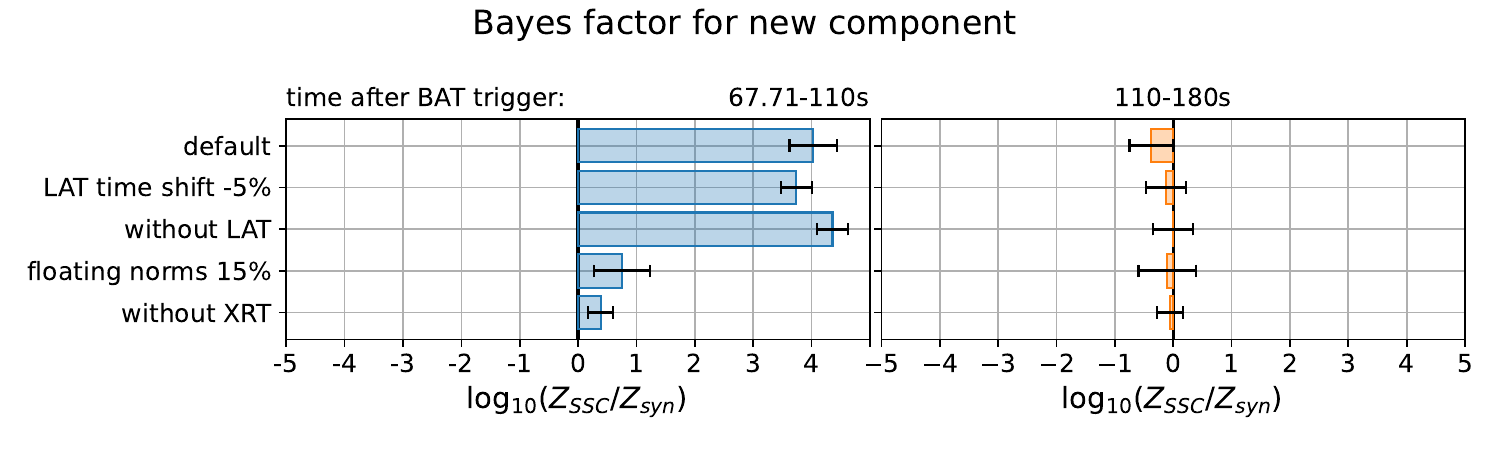}
    \caption{Bayes factor for new component for the default fit configuration and 4 perturbed cases (see text for details) for the time bins 67.71-110s and 110s-180s.}
    \label{fig:evidences}
\end{figure*}

For our default fit case, we find that for the first time bin, a Bayes factor, $\log_{10}(Z_{\rm SSC}/Z_{\rm syn})$, of $4$ is obtained. This indicates a preference for the SSC model over the synchrotron only model (see Fig.~\ref{fig:evidences}). We further find that in the second time bin this Bayes factor is essentially~$0$. We interpret these results as strong indication for the preference of the SSC over the synchrotron model in the first time bin, and no preference between the models in the second time bin.

To investigate the stability of this result, we define 4 perturbations to the default fit configuration: (1) \textit{LAT time shift} $-5\%$, where we shifted the time selection window for {\lat} photons by $-5\%$ (corresponding to 2.1~s), (2) \textit{without LAT}, where we excluded the {\lat} data, (3) \textit{floating norms} $15\%$, where we allowed for an effective area correction (floating norm to account for cross-calibration uncertainties between the different detectors) of up to $\pm15\%$ for each instrument and (4) \textit{without XRT}, where we excluded {\xrt}. A summary of our findings from these perturbation studies is shown in Fig.~\ref{fig:evidences}.

The first two cases (1 and 2) probe the contribution of {\lat} data to the fit. The shift of the time bin window (1) is motivated by the particular distribution of photons around the edges of the first time bin window: one photon 0.57~s before the start of the first time bin, and two photons 0.1~s before the end of the first time bin (see Fig.~\ref{fig:LATphotons} in the appendix). However, we find that the time shift (case (1)) has no significant effect on the preference for a new component in both time bins. Since even excluding the {\lat} data completely does not change the preference, we conclude that the few {\lat} photons bring no strong contribution to the model comparison.

In cases (3) and (4) we find that either a small systematic uncertainty in the flux or the exclusion of the {\xrt} data removes the significant preference for the SSC model in the first time bin ($\log_{10}(Z_{\rm SSC}/Z_{\rm syn})<1$, see Fig.~\ref{fig:evidences}). In the second time bin no preference is found.
In our interpretation the preference for the SSC model in the first time bin is driven by an increased {\xrt} flux compared to {\bat} and {\gbm}. Over 9 orders of magnitude in energy this small softening in the {\bat}--{\gbm} regime would grow to a significant flux suppression at the highest energies, which in turn would demand a new component in the VHE regime.

\section{Discussion and conclusions}
\label{sec:discussion}

Prior to 2018, the most constraining spectral data for probing the onset of a new spectral component came from {\xrt}, Nuclear Spectroscopic Telescope Array (NuSTAR), and {\lat}, observations of GRB~130427A \citep{2013ApJ...779L...1K,2014Sci...343...42A}. This GRB was the brightest afterglow that {\lat} had observed in the proceeding decade since its launch in 2008. No evidence for the onset of a new spectral component was found for this GRB (and for others). Post 2018, the VHE gamma-ray detections of GRB afterglow emission now allow the nature of this emission to be critically addressed. 

Observationally, for all VHE detected GRB, the apparent similarity in the level of the energy flux from the X-ray to VHE band over a broad range of energies and time periods is striking \citep{HESS_180720B,MAGIC_newComp,HESS_190829A}.
Although on energetic grounds it seems challenging for this emission to be synchrotron in origin, such a model can be realised  via a multi-zone description, which itself seems a natural extension of the standard model to consider \citep{Khangulyan+21}. 

Motivated to look deeper into broadband multi-wavelength analysis for GRB~190114C, we carried out a counts level fit of both synchrotron and SSC models, for both the first (67.71-110s) and second (110-180s) time bin in the early afterglow phase of the GRB. Our \textit{default} analysis results for GRB~190114C, not taking into account systematics, do indeed indicate a preference for the SSC description over the (single component) synchrotron description for the first time bin (67.71-110s), consistent with the results reported in \cite{MAGIC_newComp}.
However, no such preference is found for the second time bin (110-180s). Furthermore, we find that this preference for one model in the first time bin data is not stable. Our stability  analysis of the strength of preference for the SSC model over the synchrotron model finds that such a preference disappears once systematics between instruments are taken into account. Additionally, we find that the {\lat} data set does not play a relevant role in this model comparison. Instead, the normalisation of the {\xrt} data set is driving the preference. It is worth mentioning that we cannot exclude the possibility of the decaying X-ray prompt component still contributing in the first time bin (see fig.~2 in \cite{FermiNSwift} and fig.5 in \cite{AGILE_KW_GRB190114C}).

We conclude that no stable preference for either the SSC model or the synchrotron only model can be drawn from the data in the two time bins. We emphasize that this result is still compatible with the {\hess} findings for GRB~190829A \citep{Zhi_Qiu_190829A}. 

We expect these results to be general, given that GRB~190114C does not appear exceptional in terms of spectral shape.
At energies above the {\xrt} energy range, a photon index of $\sim 2$ was consistently observed by all instruments (see Table~\ref{tab:xspec} of the appendix), in agreement with the extended-time-window ensemble index seen by {\lat} \citep{Fermi_catalogue2}.
Within the {\xrt} energy range, a harder photon index of around $\sim 1.6$ was observed, also consistent with the ensemble average afterglow index for {\xrt} for the subset jointly detected by {\lat} \citep{FermiSwift_2018}.

Our results here motivate that similar investigations be carried out in the future for other bright and local GRBs with good multi-wavelength datasets. In particular, a large number of such GRBs are anticipated to be detected by the next-generation VHE telescope, the Cherenkov Telescope Array (CTA).

\section*{Acknowledgements}

The authors acknowledge support from DESY (Zeuthen, Germany), a member of the Helmholtz Association HGF. The authors would also like to thank Phil Evans, Andy Beardmore, and Kim Page for helpful input on the {\xrt}. This work was supported by the International Helmholtz-Weizmann Research School for Multimessenger Astronomy, largely funded through the Initiative and Networking Fund of the Helmholtz Association.

%%%%%%%%%%%%%%%%%%%%%%%%%%%%%%%%%%%%%%%%%%%%%%%%%%
\section*{Data Availability}

This analysis used the publicly available data from instruments on board of the \textit{Swift} and \textit{Fermi} satellites, which has been processed with the publicly available software developed by the corresponding collaborations. Additionally the data points from \citet{MAGIC_newComp} have been used. The fits have been performed with the publicly available Multi-Mission Maximum Likelihood framework \citep[3ML;][]{3ML_ref} and the UltraNest package \citep[][]{Ultranest_ref}. The code of the authors will be made available upon request.



\bibliographystyle{mnras}
\bibliography{references} % if your bibtex file is called example.bib

%\begin{thebibliography}{99}
%\end{thebibliography}

%%%%%%%%%%%%%%%%%%%%%%%%%%%%%%%%%%%%%%%%%%%%%%%%%%

%%%%%%%%%%%%%%%%% APPENDICES %%%%%%%%%%%%%%%%%%%%%

\appendix

\section{Electron Broken Power-Law Description}
\label{ap:model}

Following \citet{Jones65_ICelcool} we here demonstrate the negligible impact of inverse Compton scattering on the cooled electron spectrum. We adopt here a representative set of parameters: ${t_\mathrm{obs}=100~\mathrm{s}}$, ${E\st{iso}=3.5\times10^{53}~\mathrm{erg}}$, ${n\st{up}=1 {\rm cm}^{-3}}$, ${N_{\rm IC}=1}$, ${\varepsilon_B = 10^{-3.5}}$, ${\eta=1}$, ${p=2}$, ${\emin=10^{9}~\mathrm{eV}}$. This set of parameters fixes $\varepsilon_e$ to a value of $\varepsilon_e \approx 10^{-1.5}$.

\begin{figure}
    \centering
    \includegraphics[width=0.8\linewidth]{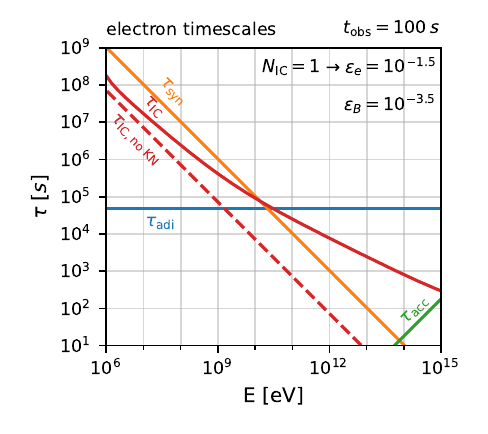}
    \includegraphics[width=0.8\linewidth]{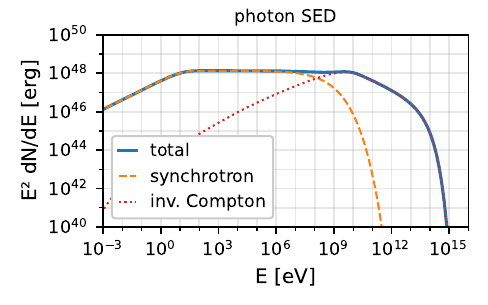}
    \caption{Above: Electron cooling timescales as a function of electron energy. Below: comoving photon energy spectrum produced through electron energy loss processes. See appendix \ref{ap:model} and section \ref{sec:model} for definitions of curves.}
    \label{fig:icCooling}
\end{figure}

The upper panel of fig.~\ref{fig:icCooling} compares the obtained inverse Compton cooling timescale $\tau_\mathrm{IC}$ (red) with the other relevant cooling timescales for the electrons (see equations \ref{eq:tadi}, \ref{eq:tsyn} and \ref{eq:tacc}). $\tau_\mathrm{IC}$ reflects an inversion of the synchrotron photon population (see lower panel of fig.~\ref{fig:icCooling}), taking into account the Klein-Nishina suppression. As can be seen from the upper panel, the role of inverse Compton cooling can be safely neglected, justifying the smooth broken power-law description.

The lower panel of fig.~\ref{fig:icCooling} shows the subsequent (comoving) spectral energy distribution produced for the example case we consider here. As expected, the inverse Compton peak sits at the same energy flux level as the synchrotron peak. Indeed, by fixing the $N_{\rm IC}$ parameter to $N_{\rm IC}=1$, the meeting of the $\tau_{\rm IC}$ and $\tau_{\rm syn}$ curves is forced to occur at approximately the break energy scale (dictated by the energy when $\tau_{\rm adi}=\tau_{\rm syn}$) in the upper panel of fig.~\ref{fig:icCooling}.

Additionally, it should be emphasised that only a small fraction of the upstream ram pressure is expected to be converted into non-thermal electrons at the shock. Therefore values $\varepsilon_e \approx 1$ are incompatible with this assumption, and would lead to a radical change to the shock jump dynamics.  
We furthermore remark, that for this case of $\varepsilon_e=10^{-1.5}$ the maximum inverse Compton flux is already similar to the synchrotron one, such that this situation represents $N_\mathrm{IC}=1$ motivated by the observations.

\section{Details on data reduction}
\label{ap:data}

For the \emph{Swift} data ({\bat} and {\xrt}), we adopt the same data selection as used in \cite{FermiNSwift}. Relevant in the {\xrt} energy range ($\lesssim$ 10 keV), we considered galactic  (``TBabs'') and extragalactic (``zTBabs'') absorption models, where the abundance and cross-section are set to ``wilm'' \citep{Wilms2000} and ``vern'' \citep{Verner1996}, respectively. Since {\xrt} is an imaging photon counter, an  on-source and an off-source (background) region of the image is used to determine the on-/off-source counts, which both follow Poissonian statistics and are included using the {\tt C-stat} \citep{cash1979}. {\bat} uses a coded mask to fit the photon counts and we include this fit result with its Gaussian uncertainty with the standard $\chi^2$ (also called S-statistic).

For {\gbm}, we used the data from two NaIs (n4 and n7; 52~keV to 875~keV) and one BGO (b0; 288~keV to 36.3~MeV). Note that all NaI channels below 50~keV are excluded due to the improper instrument response \citep{FermiNSwift}. The background for each {\gbm} detector is estimated from a first-order polynomial fit before and after GRB190114C (-135s to -35s, 260s to 375s), which shows $\chi^2 \lesssim 1$ in all energy channels. The signal (on-source) follows thus Poisson statistics, while the fitted background is included using Gaussian statistics ({\tt PG-stat}).

\begin{figure*}
    \centering
    \includegraphics[width=0.7\linewidth]{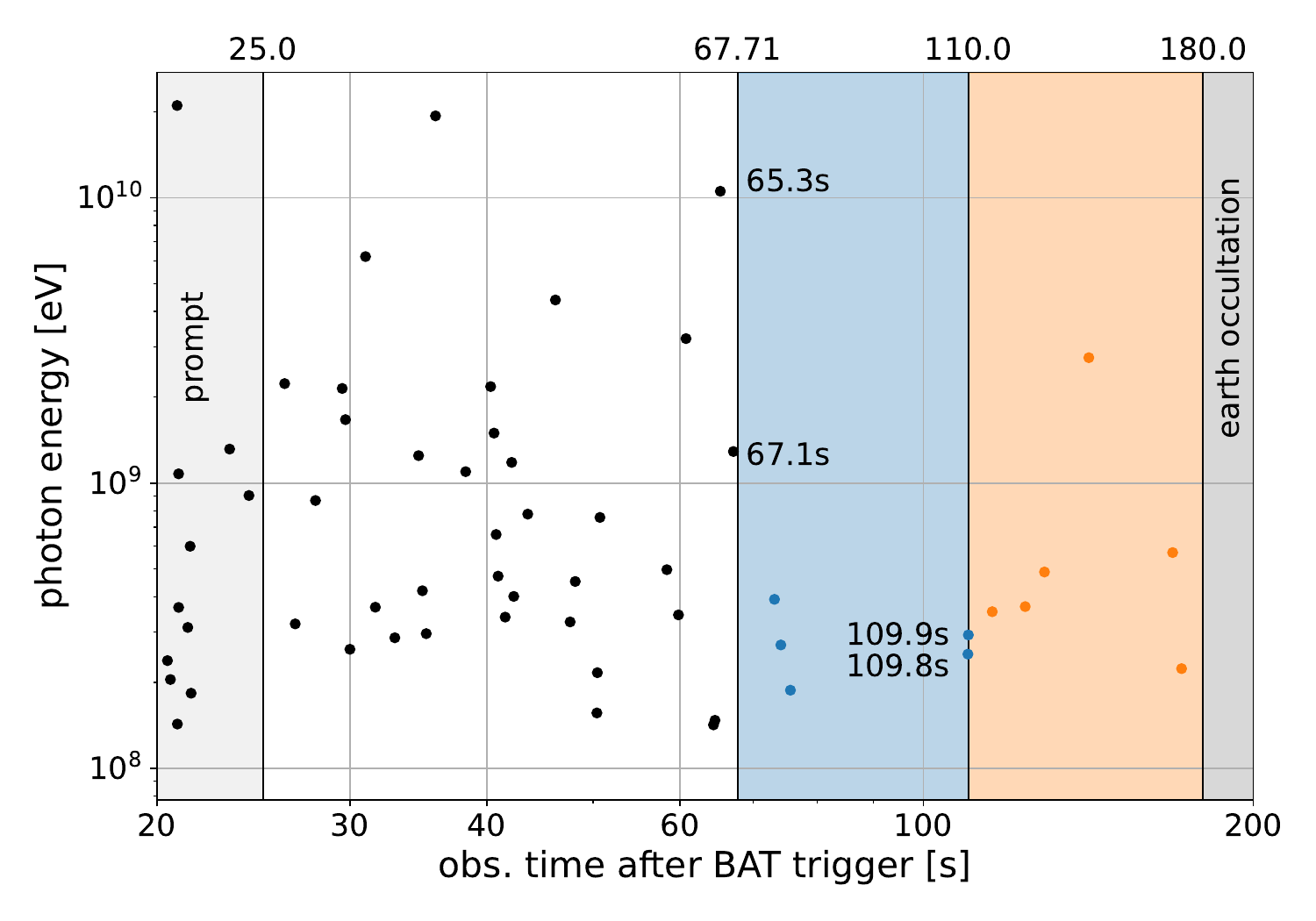}
    \caption{{\lat} photons after selection criteria as discussed in the text. The complex variability of the prompt phase ends at around 25s \citep{MAGIC_newComp} and after 180s {\lat} field of view is occulted by the earth. The blue shaded region from 67.71 to 110s and the orange one from 110 to 180s are the two time intervals with overlapping data from all 5 instruments included in this work. Note the peculiar time binning at the edges of the blue bin: a 1.28 GeV photon just about 0.6s and a 10.5 GeV photon 2.4s before the first time interval, as well as two photons of roughly 0.3 GeV within around 0.1s before the transition to the second time bin.}
    \label{fig:LATphotons}
\end{figure*}

For {\lat} data, we select {\tt P8R3Transient020E} class events from 100 MeV to 10 GeV and then apply cuts with the region of interest (RoI) of 10$^\circ$ radius centered on GRB190114C and the maximum zenith angle of 110$^\circ$. Fig.~\ref{fig:LATphotons} shows the {\lat}--detected events passing the selections. For the galactic and isotropic diffuse backgrounds, we used the latest templates provided by the {\lat} collaboration, {\tt gll\_iem\_v07} and {\tt iso\_P8R3\_TRANSIENT020E\_V3\_v1}, respectively. 

Before performing the joint-fit spectral analysis, we fit the individual data and cross-check with \cite{FermiNSwift}. The results agree within 1$\sigma$ level, see Table~\ref{tab:xspec}.

\setcounter{table}{0}

\begin{table*}
\centering
\begin{tabular}{|l|l|l|l|}
\hline
dataset                      & time bin & XSPEC                        & 3ML                \\ \hline \hline
\multirow{2}{*}{{\xrt}}         & 1             &  $-1.649^{+0.068}_{-0.069}$  & $-1.75 \pm 0.07$   \\ \cline{2-4} 
                             & 2             & $-1.611^{+0.057}_{-0.058}$   & $-1.68 \pm 0.06$   \\ \hline \hline
\multirow{2}{*}{{\bat}}         & 1             & $-2.009^{+0.036}_{-0.036}$   & $-2.009 \pm 0.027$ \\ \cline{2-4} 
                             & 2             & $-2.021^{+0.040}_{-0.041}$   & $-2.001 \pm 0.032$ \\ \hline \hline
\multirow{2}{*}{{\gbm}}         & 1             & $-1.989^{+0.093}_{-0.101}$   & $-1.94 \pm 0.10$   \\ \cline{2-4} 
                             & 2             & $-1.822^{+0.118}_{-0.138}$   & $-1.81 \pm 0.14$   \\ \hline \hline
\multirow{2}{*}{{\bat}+{\gbm}}     & 1             & $-2.007^{+0.034}_{-0.034}$   & $-2.004 \pm 0.027$ \\ \cline{2-4} 
                             & 2             & $-2.007^{+0.039}_{-0.039}$   & $-1.995 \pm 0.032$ \\ \hline \hline
\multirow{2}{*}{{\bat}+{\gbm}+{\lat}} & 1             & $-2.027^{+0.027}_{-0.028}$   & $-2.036 \pm 0.023$ \\ \cline{2-4} 
                             & 2             & $-2.017^{+0.030}_{-0.031}$   & $-2.013 \pm 0.026$ \\ \hline
\end{tabular}
\caption{Validation of 3ML results against XSPEC: We compare the spectral index of a single power law fit for time bins 1 (67.71-110s after $T_{0, \mathrm{BAT}}$) and 2 (110-180s after $T_{0, \mathrm{BAT}}$). Results agree within $1\sigma$-level.}
\label{tab:xspec}
\end{table*}

\section{Details on joint fitting procedure and model comparison}
\label{ap:bayes}

We utilize a {\tt python} package, the Multi-Mission Maximum Likelihood framework \citep[3ML;][]{3ML_ref}, for our analysis.

We define $\mathcal{L}(\mathcal{D}|\vec{\theta}, \mathcal{M})$ as the probability to obtain a sample of data ($\mathcal{D}$) conditioning for a set of parameters ($\vec{\theta}$) and explicitly also the model ($\mathcal{M}$). Following a Bayesian approach, we define $\pi(\vec{\theta})$ as the prior probability distribution of $\vec{\theta}$. We assume uniform ($p\in[1.5, 2.5]$) and log-uniform ($\varepsilon_B\in[10^{-5}, 1]$, $\eta\in[10^{-3}, 10^{5}]$, $F\st{syn}\in[10^{-9}, 10^{-7}]$ and $N\st{IC}\in[10^{-2}, 10]$) priors. We derive the posterior probability distribution and the evidence $Z$ \citep[e.g.][]{KassRaftery95,Trotta2008}
\begin{eqnarray}
    Z = \int \dd \vec{\theta}  \; \mathcal{L}(\mathcal{D}|\vec{\theta},\mathcal{M}) \: \pi(\vec{\theta},\mathcal{M}) \:,
\end{eqnarray}
with the nested sampling Monte Carlo algorithm MLFriends, implemented in the {\tt python} package UltraNest \citep{Ultranest_ref}. 
$Z$ intuitively represents the posterior averaged over the entire parameter space. It is of particular interest for our case of model comparison, since it allows for a quantitative measure of our preference of model $\mathcal{M}_1$ over $\mathcal{M}_2$ via the Bayes factor $B_{12}=Z_1/Z_2$.

\begin{figure*}
    \centering
    \includegraphics[width=\linewidth]{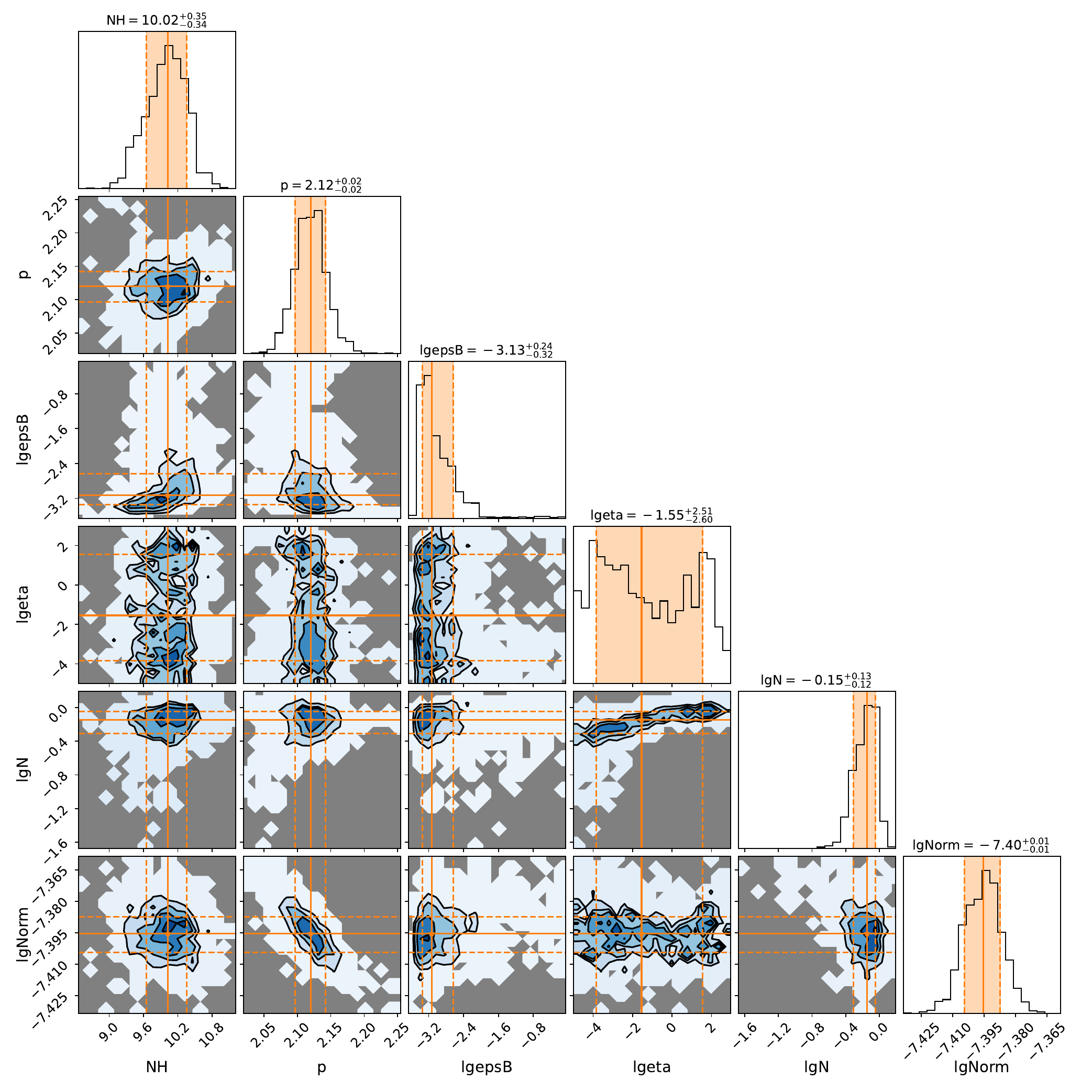}
    \caption{Posterior distributions for SSC default fit for the first time interval $67.71-110$s. NH is the column density of the photoelectric absorption in $10^{22}$~atoms~per~cm$^2$, $p$ is the spectral index of the uncooled electrons, $\varepsilon_B = 10^\mathrm{lgepsB}$ defines the magnetic field strength, $\eta= 10^\mathrm{lgeta}$ is inversely proportional to the maximum energy of the electrons and thus also photons, $N\st{IC}= 10^\mathrm{lgN}$ is the relative height of the inverse Compton component to the syncchrotron component and $ F\st{syn}= 10^\mathrm{lgNorm}$ is the reference value for the synchrotron photon flux in $erg/(cm^2s)$ at $\Eg\ut{obs} = 100$~keV.}
    \label{fig:corner_SSC_default_t0}
\end{figure*}

\begin{figure*}
    \centering
    \includegraphics[width=\linewidth]{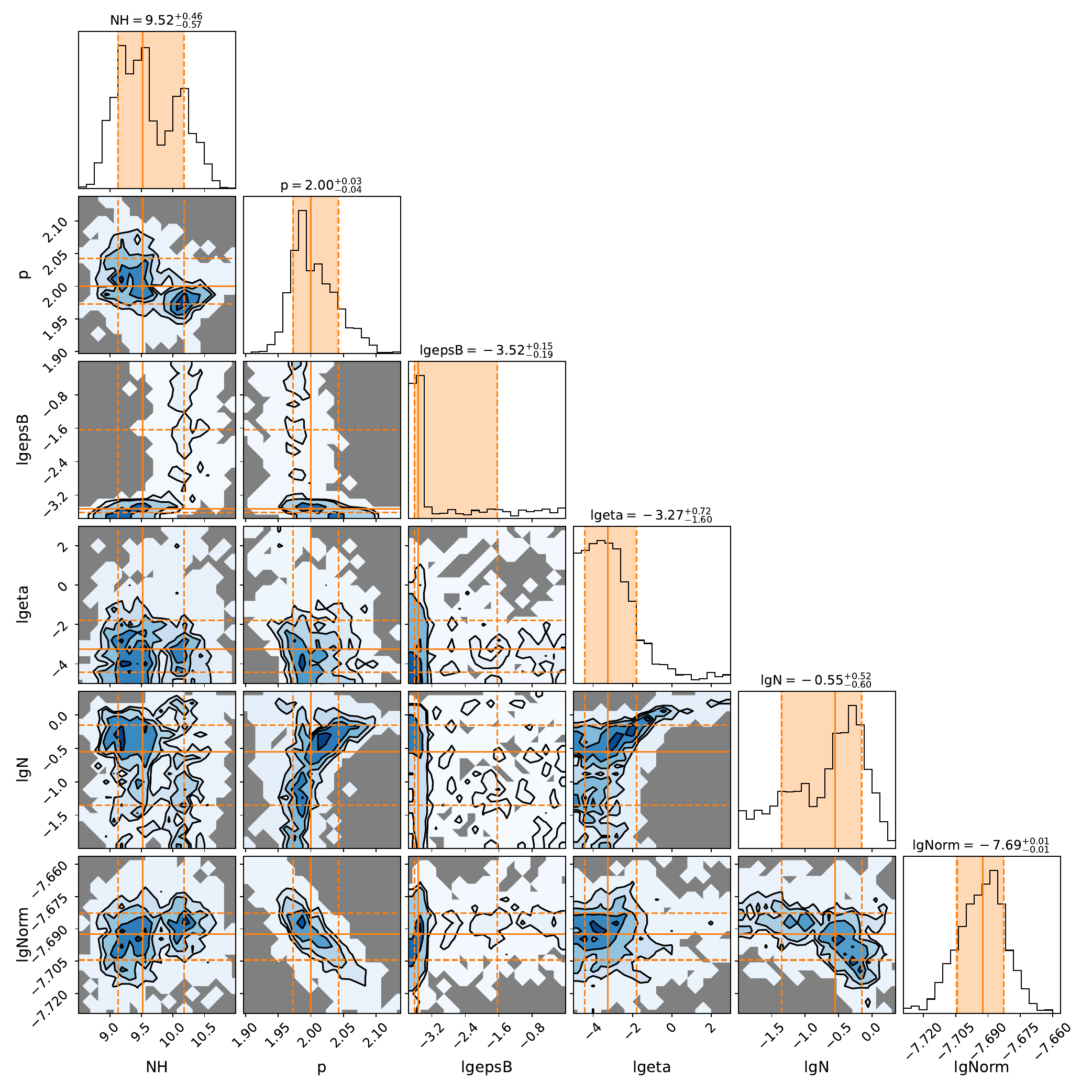}
    \caption{Posterior distributions for SSC default fit for the second time interval $110-180$s. Notation as in Fig.~\ref{fig:corner_SSC_default_t0}}
    \label{fig:corner_SSC_default_t1}
\end{figure*}

\begin{figure*}
    \centering
    \includegraphics[width=\linewidth]{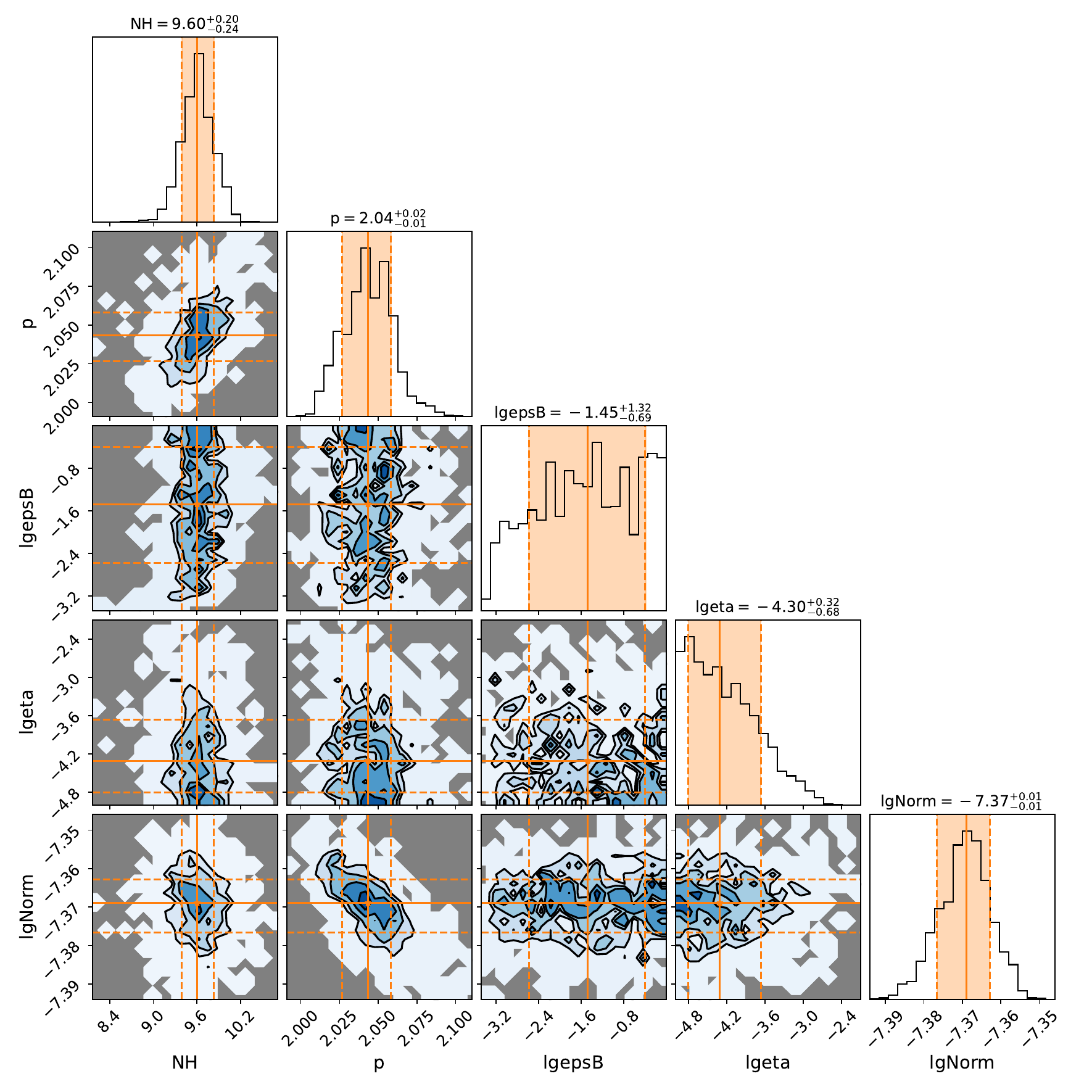}
    \caption{Posterior distributions for syn. only ($N\st{IC}= 0$) default fit for the first time interval $67.71-110$s. Notation as in Fig.~\ref{fig:corner_SSC_default_t0}}
    \label{fig:corner_syn_default_t0}
\end{figure*}

\begin{figure*}
    \centering
    \includegraphics[width=\linewidth]{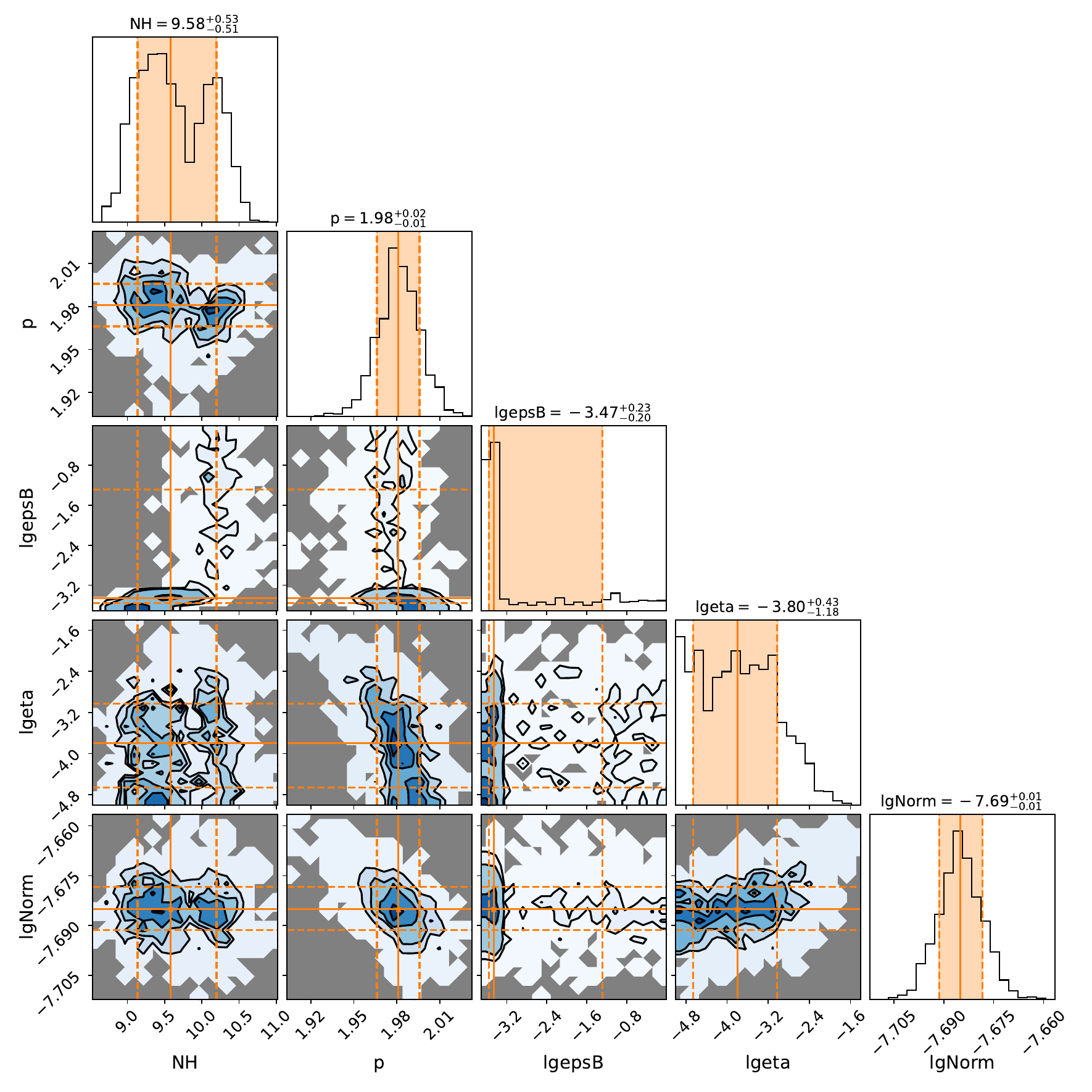}
    \caption{Posterior distributions for syn. only ($N\st{IC}= 0$) default fit for the second time interval $110-180$s. Notation as in Fig.~\ref{fig:corner_SSC_default_t0}}
    \label{fig:corner_syn_default_t1}
\end{figure*}

In Figs.~\ref{fig:corner_SSC_default_t0},~\ref{fig:corner_SSC_default_t1},~\ref{fig:corner_syn_default_t0} and \ref{fig:corner_syn_default_t1} we show the posterior distributions obtained for the default case for both time bins.

%%%%%%%%%%%%%%%%%%%%%%%%%%%%%%%%%%%%%%%%%%%%%%%%%%

% Don't change these lines
\bsp	% typesetting comment
\label{lastpage}
\end{document}